\documentstyle[12pt]{article}

\begin{document}

\pagenumbering{roman} 

%begin \large
 
%{\large

%\textcolor{Black}{
\begin{center}
{\em Statistical properties of the quantum anharmonic oscillator.}
\end{center}
%}

\begin{center}
{\tt Maciej M. Duras}
\end{center}

\begin{center}
Institute of Physics, Cracow University of Technology, 
ulica Podchor\c{a}\.zych 1, PL-30084 Cracow, Poland.
\end{center}

\begin{center}
Email: mduras @ riad.usk.pk.edu.pl
\end{center}

\begin{center}
``Polymorphism in Condensed Matter
International workshop'';
November 13th, 2006 -  Novemer 17th, 2006;
Max Planck Institute for the Physics of Complex Systems,
Dresden, Germany (2006).
\end{center}

\begin{center}
AD 2006 November 15th
\end{center}

{\tt I. Abstract}

The random matrix ensembles (RME) of Hamiltonian matrices, 
e.g. Gaussian random matrix ensembles (GRME) 
and Ginibre random matrix ensembles (Ginibre RME), 
are applicable to following quantum statistical systems: nuclear systems, 
molecular systems, 
condensed phase systems, disordered systems, 
and two-dimensional electron systems (Wigner-Dyson electrostatic analogy). 
A family of quantum anharmonic oscillators is studied 
and the numerical investigation of their eigenenergies is presented. 
The statistical properties of the calculated eigenenergies are compared 
with the theoretical predictions inferred from the random matrix theory. 
Conclusions are derived.

{\tt II. Quantum harmonic oscillator in $D=1$ spatial dimension.}

{\sl Firstly:} Let us consider the Hilbert space:
%\textcolor{Black}{
\begin{equation}
{\cal V}_1 = L^2({\bf R}, {\bf C}, {\rm d}x),
\label{Hilbert-space-L2-D=1}
\end{equation}
%}
of the complex valued wave functions $\Psi$ that are (modulus) square integrable
on the set ${\bf R}$ of the real numbers,
and ${\bf C}$ is the set of the complex numbers.
The Hilbert space ${\cal V}_1$ is separable space,
and its orthonormal basis ${\cal B}$ is a set of Hermite's functions $\Psi_r$ 
(Fok's functions, eigenfunctions of the dimensionless 
Hamiltonian operator $\hat{{\cal H}}$ 
of the quantum harmonic oscillator in $D=1$ spatial dimension):
%\textcolor{Black}{
\begin{equation}
\Psi_r(x) = N_r H_r(x) \exp(-\frac{1}{2}x^2), 
N_r = [ \sqrt{\pi} r! 2^r ]^{\frac{1}{2}}, r \in {\bf N},
\label{Hermite-function-definition}
\end{equation}
%}
where ${\bf N}$ is a set of natural numbers including zero,
whereas:
%\textcolor{Black}{
\begin{equation}
H_r(x) = (-1)^r \exp(x^2) \frac{{\rm d}^r}{{\rm d}x^r} \exp(-x^2), 
\label{Hermite-polynomial-definition}
\end{equation}
%}
is $r$th Hermite's polynomial.
We assume from now that all the operators are dimensionless (nondimensional).
The dimensionless quantum operator $\hat{x}$ of the $x$th component 
of the position vector (radius vector) 
acts on the  Hermite's basis function as follows:
%\textcolor{Black}{
\begin{equation}
\hat{x} \Psi_r = \sqrt{\frac{r}{2}} \Psi_{r-1} 
+ \sqrt{\frac{r+1}{2}} \Psi_{r+1}, 
\label{x-operator-Hermite-function-action}
\end{equation} 
%}
whereas the quantum operator $\hat{p}_x$ of the $x$th component 
of linear momentum vector 
is given in the basis by the following formula:
%\textcolor{Black}{
\begin{equation}
\hat{p}_x \Psi_r = \frac{1}{i} \sqrt{\frac{r}{2}} \Psi_{r-1} 
- \frac{1}{i} \sqrt{\frac{r+1}{2}} \Psi_{r+1}. 
\label{p_x-operator-Hermite-function-action}
\end{equation}  
%}
Matrix elements $x_{l,r}$ and $(p_{x})_{l,r}$ 
of these operators equal correspondingly:
%\textcolor{Black}{
\begin{equation}
x_{l,r} = \langle \Psi_l | \hat{x} \Psi_r \rangle = 
\sqrt{\frac{r}{2}} \delta_{l,r-1} 
+ \sqrt{\frac{r+1}{2}} \delta_{l,r+1}, 
\label{x-operator-Hermite-function-matrix-element}
\end{equation} 
%}
and
%\textcolor{Black}{
\begin{equation}
(p_{x})_{l,r} = \langle \Psi_l | \hat{p}_x \Psi_r \rangle = 
\frac{1}{i} \sqrt{\frac{r}{2}} \delta_{l,r-1} 
- \frac{1}{i} \sqrt{\frac{r+1}{2}} \delta_{l,r+1}, 
\label{p_x-operator-Hermite-function-matrix-element}
\end{equation}  
%}
where
%\textcolor{Black}{
\begin{equation}
\delta_{l,r} = 
\left\{ 
\begin{array}{cc}
1, & l=r \\
0, & l \neq r \\
\end{array}
\right. , 
\label{Kronecker-delta-definition}
\end{equation}
%}
is discrete Kronecker's delta (it is not continuous Dirac's delta).
The quantum operator $\hat{x}^2$ of the square of the $x$th component of radius vector 
after acting on the basis function reads:
%\textcolor{Black}{
\begin{equation}
\hat{x}^2 \Psi_r = 
\frac{1}{2} 
[ \sqrt{r-1} \sqrt{r} \Psi_{r-2}
+ (2r + 1) \Psi_{r} 
+ \sqrt{r+1} \sqrt{r+2} \Psi_{r+2} ], 
\label{x^2-operator-Hermite-function-action}
\end{equation}
%}
and its matrix element $(x^2)_{l,r}$ is equal:
%\textcolor{Black}{
\begin{equation}
(x^2)_{l,r} = 
\frac{1}{2} 
[ \sqrt{r-1} \sqrt{r} \delta_{l,r-2}
+ (2r + 1) \delta_{l,r} 
+ \sqrt{r+1} \sqrt{r+2} \delta_{l,r+2} ], 
\label{x^2-operator-Hermite-function-matrix-element}
\end{equation} 
%}
whereas the quantum operator $\hat{p}_x^2$ of the square 
of the $x$th component of the linear momentum vector
has the following representation in the basis:
%\textcolor{Black}{
\begin{equation}
\hat{p}_x^2 \Psi_r = 
\frac{1}{2} 
[ -\sqrt{r-1} \sqrt{r} \Psi_{r-2}
+ (2r + 1) \Psi_{r} 
- \sqrt{r+1} \sqrt{r+2} \Psi_{r+2} ], 
\label{p^2-operator-Hermite-function-action}
\end{equation}
%}
and its matrix element $(p_x^2)_{l,r}$ is equal:
%\textcolor{Black}{
\begin{equation}
(p_x^2)_{l,r} = 
\frac{1}{2} 
[ - \sqrt{r-1} \sqrt{r} \delta_{l,r-2}
+ (2r + 1) \delta_{l,r} 
- \sqrt{r+1} \sqrt{r+2} \delta_{l,r+2} ]. 
\label{p^2-operator-Hermite-function-matrix-element}
\end{equation} 
%}
Therefore, {\sl the dimensionless (nondimensional)} 
quantum Hamiltonian operator $\hat{{\cal H}}$ reads:
%\textcolor{Black}{
\begin{equation}
\hat{{\cal H}} = \hat{p}_x^2 + \hat{x}^2, 
\label{H-operator-harmonic-1D-definition}
\end{equation}  
%}
acting on the basis function, it produces:
%\textcolor{Black}{
\begin{equation}
\hat{{\cal H}} \Psi_r = 
(2r + 1) \Psi_{r}, 
\label{H-operator-harmonic-1D-Hermite-function-action}
\end{equation}
%}
and its matrix element ${\cal H}_{l,r}$ is equal:
%\textcolor{Black}{
\begin{equation}
{\cal H}_{l,r} = 
(2r + 1) \delta_{l,r}, 
\label{H-operator-harmonic-1D-Hermite-function-matrix-element}
\end{equation} 
where
\begin{equation}
\epsilon_r = 
2r + 1, 
\label{H-operator-harmonic-1D-Hermite-function-eigenvalue}
\end{equation} 
%}
is the $r$th eigenenergy of $\hat{{\cal H}}$.
The eigenenergies are simply all odd natural numbers, 
and the quantum Hamiltonian is diagonal operator, 
and its matrix representation is diagonal $\infty \times \infty$ matrix.
Note, that if one introduces the notion 
of nearest neighbour energy spacing (NNS):
%\textcolor{Black}{
\begin{equation}
s_r = \epsilon_{r+1} - \epsilon_r, r=1, \cdots (N-1), N \geq 2,
\label{nearest-neighbour-spacing-definition}
\end{equation} 
%}
then for the quantum harmonic oscillator it holds:
%\textcolor{Black}{
\begin{equation}
s_r = 2 = {\rm const}, 
\label{nearest-neighbour-spacing-1DIM_quantum_harmonic_oscillator}
\end{equation} 
%}
so the eigenenergies are equidistant. 
Here $N$ is the number of energies dealt with 
(it is also a dimension of
truncated Hilbert space ${\cal V}_{1,N} \equiv {\bf C}^N$). 
The eigenfunctions of ${\cal V}_{1,N}$
are $N$-component constant complex vectors (analogs of spinors),
and the operators acting on it are $N \times N$ deterministic complex-valued matrices.
The probability distribution $P_{N-1}$ of the spacings
is discrete one point distribution for any finite value of $N, N \geq 2$: 
%\textcolor{Black}{
\begin{equation}
P_{N-1}(s) = \frac{1}{N-1} \delta_{s,2},
\label{nearest-neighbour-spacing-distribution-finite-N}
\end{equation} 
%}
tending in the thermodynamical limit $N \rightarrow \infty$
to the singular Dirac's delta distribution: 
%\textcolor{Black}{
\begin{equation}
P_\infty(s)=\delta(s-2). 
\label{nearest-neighbour-spacing-distribution-infinite-N}
\end{equation} 
%}

{\sl Secondly,} let us perform more difficult task 
consisting of calculating all the absolute moments:
%\textcolor{Black}{
\begin{equation}
(m_n)_{l,r}=m_n=(x^n)_{l,r} = \langle \Psi_l | \hat{x}^n \Psi_r \rangle 
= \int_{-\infty}^{\infty} \Psi_l^{\star}(x) x^n \Psi_r(x){\rm d}x, 
\label{x-operator-nth-abolute-moment}
\end{equation} 
%}
of the position operator $\hat{x}$ in the Hermite's (Fok's) basis ${\cal B}$. 
One can calculate the lower moments manually, {\sl e. g.}, 
using recurrence relations, matrix algebra, {\sl etc.},
but it is tedious (even for $3 \leq n \leq 6$). 
If one wants to calculate {\sl all} the moments then he must return 
to the beautiful XIX century mathematics methods
and after some reasoning he obtains the exact formula:
%\textcolor{Black}{
\begin{eqnarray}
& & (m_n)_{l,r}=m_n=(x^n)_{l,r}=  
\nonumber \label{x-operator-nth-abolute-moment-matrix-element-1} \\ 
& & = [1 - (-1)^{n+l+r}] \sum_{j=0}^{[l/2]} \sum_{k=0}^{[r/2]} [ (-1)^{j+k} 
\frac{\sqrt{l!}}{j! (l-2j)!} \frac{\sqrt{r!}}{k! (r-2k)!} \cdot 
\nonumber \label{x-operator-nth-abolute-moment-matrix-element-2} \\
& & \cdot 2^{\frac{l}{2}+\frac{r}{2}-2j-2k-1}
\Gamma(\frac{n+l+r-2j-2k+1}{2}) ], 
\label{x-operator-nth-abolute-moment-matrix-element-3} 
\end{eqnarray}
%}
where $[\cdot]$ is entier (step) function, $\Gamma$ is Euler's gamma function.
Therefore, the matrix representation of the 
even power operators ${\hat{x}^{2p}}$ 
in the basis ${\cal B}$
are hermitean (symmetrical real) matrices with nonzero diagonal 
and nonzero $p$ subdiagonals 
(and nonzero $p$ superdiagonals), 
where the distance of the nearest superdiagonals (or subdiagonals) is 2 
(the diagonal is also distant by 2 from the nearest super- and sub-diagonal),
whereas the odd power operators ${\hat{x}^{2p+1}}$ in the basis ${\cal B}$
are hermitean (symmetrical real) matrices with zero diagonal 
and $p$ nonzero subdiagonals 
(and $p$ nonzero superdiagonals), 
where the distance of the nearest superdiagonals (or subdiagonals) is 2 
(the nearest super- and sub-diagonal are also distant by 2).
The physical interpretation of the superdiagonals (and subdiagonals) 
is connected with the absorption (emission) of phonons.

{\tt III. Quantum anharmonic oscillator in $D=1$ spatial \newline dimension.}

{\sl Thirdly,} we are ready to deal with the quantum anharmonic oscillator 
in $D=1$ spatial dimension.
Its dimensionless Hamiltonian operator 
$\hat{{\cal H}}_{1, {\rm anharm}}^{S}$ reads:
%\textcolor{Black}{
\begin{equation}
\hat{{\cal H}}_{1, {\rm anharm}}^{S} 
=\hat{{\cal H}}+ \sum_{s=0}^{S} a_s \hat{x}^s, 
\label{H-operator-anharmonic-1D-S}
\end{equation}
%}
where $S$ is a degree of the anharmonicity of the oscillator, 
and the prefactors $a_s$ are the strengths of anharmonicity.
The matrix elements of the anharmonic Hamiltonian operator are:
%\textcolor{Black}{
\begin{equation}
(\hat{{\cal H}}_{1, {\rm anharm}}^{S})_{l,r} 
=\epsilon_r \delta_{l,r} + \sum_{s=0}^{S} a_s (x^s)_{l,r}, 
\label{H-operator-anharmonic-1D-S-matrix-element}
\end{equation}
%}
where the representation of the quantum anharmonic oscillator 
in the quantum harmonic oscillator basis ${\cal B}$
is mathematically correct, because the basis ${\cal B}$ is a complete set, 
and the Hilbert space of the eigenfunctions
of the anharmonic oscillator is isomorphic 
to the Hilbert space ${\cal V}_1$ for the harmonic oscillator, 
provided that
the anharmonic potential:
%\textcolor{Black}{
\begin{equation}
{\cal U}_{1, {\rm anharm}}^{S}(x) = \sum_{s=0}^{S} a_s x^s, 
\label{U-anharmonic-potential-1D-S}
\end{equation}
%}
is bounded from below (there are no scattering eigenstates).

{\sl Fourthly,} we repeat the ``Bohigas conjecture'' 
that the fluctuations of the spectra of the quantum systems
that correspond clasically to the chaotic systems generally obey 
the spectra of the Gaussian random matrix ensembles.
The quantum integrable systems correspond to the classical integrable systems 
in the semiclassical limit
\cite{Bohigas 1984,Ozorio de Almeida 1988}.
The probability distributions $P_\beta$ of the nearest neighbour spacing 
for the Gaussian orthogonal ensemble GOE(2) of $2 \times 2$ Gaussian distributed
real-valued  symmetric random matrix variables ($\beta=1)$,
for the Gaussian unitary ensemble GUE(2) of $2 \times 2$ Gaussian distributed
complex-valued hermitean random matrix variables ($\beta=2)$,   
for the Gaussian symplectic ensemble GSE(2) of $2 \times 2$ Gaussian distributed
quaternion-valued selfdual hermitean random matrix variables ($\beta=4$),
and for the Poisson ensemble (PE) of the random diagonal matrices 
with homogeneously distributed eigenvalues
on the real axis ${\bf R}$ are given by the formulae:
%\textcolor{Black}{
\begin{equation}
P_\beta(s) = \theta(s) A_{\beta} s^{\beta} \exp(-B_{\beta} s^2), 
\label{P_beta-NNS-distribution-GOEGUEGSE} 
\end{equation} 
%}
for the Gaussian ensembles, 
and
%\textcolor{Black}{ 
\begin{equation}
P_0(s) = \theta(s) \exp(- s), 
\label{P_beta-NNS-distribution-PE} 
\end{equation} 
%}
for the Poisson ensemble,
where $\theta$ is Heaviside's unit step function   
\cite{Haake 1990,Guhr 1998,Mehta 1990 0,Reichl 1992,Bohigas 1991,Porter 1965,Brody 1981,Beenakker 1997,Ginibre 1965,Mehta 1990 1}. 
The constant numbers:
%\textcolor{Black}{
\begin{equation}
A_\beta = 2 \frac{\Gamma^{\beta+1}((\beta+2)/2)}
                 {\Gamma^{\beta+2}((\beta+1)/2)}
\qquad {\rm and} \qquad 
B_\beta = \frac{\Gamma^2((\beta+2)/2)}
               {\Gamma^2((\beta+1)/2)},
\label{A_beta-B_beta-constants}
\end{equation}
%}
are given by the formulae: $A_1=\pi/2$, $B_1=\pi/4$ (GOE),
$A_2=32/\pi^2$, $B_2=4/\pi$ (GUE), and $A_4=262144/729\pi^3$,
$B_4=64/9\pi$ (GSE), respectively.
For the Gaussian ensembles the energies are characterized by the ``level repulsion'' 
of degree $\beta$ near the origin
(vanishing spacing $s=0$), 
and the probability distributions $P_{\beta}$ vanish at the origin, 
and the quantum system
with ``level repulsion'' is cast to the class of quantum chaotic systems.
For the Poisson ensembles the energies are characterized 
by the ``level clustering'' near the origin
(vanishing spacing $s=0$), 
and the probability distributions $P_0$ has maximum at the origin, 
and the quantum system
with ``level clustering'' are treated as the quantum integrable system. 
After many numerical experiments conducted with different quantum anharmonic oscillators 
(up to the sextic quantum anharmonic oscillator $S=6$) 
we draw conclusion that majority of them behaves like quantum
integrable systems, the eigenenergies tend to cluster, 
the histogram of nearest neighbour spacing is closer to
the $P_0$ distribution resulting from the Poisson ensemble.

{\tt IV. Quantum harmonic oscillator in $D \geq 1$ spatial dimensions.}

{\sl Fifthly:} Quantum harmonic oscillator in $D$ spatial dimensions is a solvable analytically model.
In order to make the deliberations easier we reduce our present interest to the first quantization.
Therefore the relevant Hilbert space ${\cal V}_D$ is a $D$-dimensional tensor (Cartesian) product
of the one-dimensional Hilbert spaces  ${\cal V}_1$:
%\textcolor{Black}{
\begin{equation}
{\cal V}_D = \bigotimes_{j=1}^{D} {\cal V}_1,
\label{Hilbert-space-L2-D-tensor-product}
\end{equation}
%}  
whereas the harmonic oscillator's eigenfunction $\Psi_{\bf r}$ 
in $D$ dimensions is, neither symmetrized nor antisymmetrized, 
tensor product of the eigenfunctions in one dimension:
%\textcolor{Black}{
\begin{equation}
\Psi_{\bf r}({\bf x}) = \prod_{j=1}^{D} \Psi_{r_{j}}(x_j), {\bf x}=(x_1,...,x_D) \in {\bf R}^D, 
{\bf r}=(r_1,...,r_D) \in {\bf N}^D,
\label{eigenfunction-tensor-product-definition-D}
\end{equation}
%}
where we used boldface font for the $D$-tuples ${\bf x},$ and ${\bf r}.$
It follows that:
%\textcolor{Black}{
\begin{equation}
\Psi_{\bf r}({\bf x}) = N_{\bf r} H_{\bf r}({\bf x}) \exp(-\frac{1}{2}{\bf x}^2), 
N_{\bf r} = \prod_{j=1}^{D} N_{r_{j}},  H_{\bf r}({\bf x})= \prod_{j=1}^{D} H_{r_{j}}(x_j).
\label{eigenfunction-Hermite-function-definition-D}
\end{equation}
%}
One can also draw a conclusion that the Hilbert space:
%\textcolor{Black}{
\begin{equation}
{\cal V}_D = L^2({\bf R}^D, {\bf C}, {\rm d}x), D \geq 1,
\label{Hilbert-space-L2-D}
\end{equation}
%}
is composed of the complex valued wave functions $\Psi$ that are (modulus) square integrable
on the set ${\bf R}^D$.
The Hilbert space ${\cal V}_D$ is separable space,
and its orthonormal basis ${\cal B}_D$ is a set of Hermite's functions $\Psi_{\bf r}$ in $D$ dimensions 
(Fok's functions in $D$ dimensions, eigenfunctions of the dimensionless 
Hamiltonian operator $\hat{{\cal H}}_D$ 
of the quantum harmonic oscillator in $D$ spatial dimensions).
We assume from now that all the operators are dimensionless (nondimensional).
The dimensionless quantum operator $\hat{x}_j$ of the $j$th component 
of the position operator $\hat{{\bf x}}$ 
acts on the  Hermite's basis function as follows:
%\textcolor{Black}{
\begin{equation}
\hat{x}_j \Psi_{r_{j}} = \sqrt{\frac{r_j}{2}} \Psi_{r_j-1} 
+ \sqrt{\frac{r_j+1}{2}} \Psi_{r_j+1}, j=1,...,D, 
\label{x_j-operator-Hermite-function-action}
\end{equation} 
%}
whereas the quantum operator $\hat{p}_j$ of the $j$th component 
of linear momentum operator $\hat{{\bf p}}$
is given in the basis by the following formula:
%\textcolor{Black}{
\begin{equation}
\hat{p}_{x_{j}} \Psi_{r_{j}} = \frac{1}{i} \sqrt{\frac{r_j}{2}} \Psi_{r_j-1} 
- \frac{1}{i} \sqrt{\frac{r_j+1}{2}} \Psi_{r_j+1}. 
\label{p_x_j-operator-Hermite-function-action}
\end{equation}  
%}
Matrix elements $(x_j)_{l_j,r_j}$ and $(p_{x_{j}})_{l_j,r_j}$ 
of these operators are given by:
%\textcolor{Black}{
\begin{equation}
(x_j)_{l_j,r_j} = \langle \Psi_{l_{j}} | \hat{x}_j \Psi_{r_{j}} \rangle = 
\sqrt{\frac{r_j}{2}} \delta_{l_j,r_j-1} 
+ \sqrt{\frac{r_j+1}{2}} \delta_{l_j,r_j+1}, 
\label{x_j-operator-Hermite-function-matrix-element}
\end{equation} 
%}
and
%\textcolor{Black}{
\begin{equation}
(p_{x_{j}})_{l_j,r_j} = \langle \Psi_{l_j} | \hat{p}_{x_{j}} \Psi_{r_{j}} \rangle = 
\frac{1}{i} \sqrt{\frac{r_j}{2}} \delta_{l_j,r_j-1} 
- \frac{1}{i} \sqrt{\frac{r_j+1}{2}} \delta_{l_j,r_j+1}, 
\label{p_x_j-operator-Hermite-function-matrix-element}
\end{equation}  
%}
where
%\textcolor{Black}{
\begin{equation}
\delta_{{\bf l},{\bf r}} = \prod_{j=1}^{D} \delta_{l_j,r_j}, 
\label{Kronecker-delta-definition-D}
\end{equation}
%}
is discrete Kronecker's delta in $D$ dimensions (it is not continuous Dirac's delta).
The quantum operator $\hat{x}_j^2$ of the square of the operator $\hat{x}_j$ 
after acting on the basis function reads:
%\textcolor{Black}{
\begin{equation}
\hat{x}_j^2 \Psi_{r_{j}} = 
\frac{1}{2} 
[ \sqrt{r_j-1} \sqrt{r_j} \Psi_{r_j-2}
+ (2r_j + 1) \Psi_{r_{j}} 
+ \sqrt{r_j+1} \sqrt{r_j+2} \Psi_{r_j+2} ], 
\label{x_j^2-operator-Hermite-function-action}
\end{equation}
%}
and its matrix element $(x_j^2)_{l_j,r_j}$ reads:
%\textcolor{Black}{
\begin{equation}
(x_j^2)_{l_j,r_j} = 
\frac{1}{2} 
[ \sqrt{r_j-1} \sqrt{r_j} \delta_{l_j,r_j-2}
+ (2r_j + 1) \delta_{l_j,r_j} 
+ \sqrt{r_j+1} \sqrt{r_j+2} \delta_{l_j,r_j+2} ], 
\label{x_j^2-operator-Hermite-function-matrix-element}
\end{equation} 
%}
whereas the quantum operator $\hat{p}_{x_{j}}^2$ of the square 
of the operator $\hat{p}_{x_{j}}$ has the following representation in the basis:
%\textcolor{Black}{
\begin{equation}
\hat{p}_{x_{j}}^2 \Psi_{r_{j}} = 
\frac{1}{2} 
[ -\sqrt{r_j-1} \sqrt{r_j} \Psi_{r_j-2}
+ (2r_j + 1) \Psi_{r_{j}} 
- \sqrt{r_j+1} \sqrt{r_j+2} \Psi_{r_j+2} ], 
\label{p_x_j^2-operator-Hermite-function-action}
\end{equation}
%}
and its matrix element $(p_{x_{j}}^2)_{l_j,r_j}$ is equal:
%\textcolor{Black}{
\begin{eqnarray}
& & (p_{x_{j}}^2)_{l_j,r_j} =
\label{p_x_j^2-operator-Hermite-function-matrix-element_1} \nonumber \\ 
& & =\frac{1}{2} 
[ - \sqrt{r_j-1} \sqrt{r_j} \delta_{l_j,r_j-2}
+ (2r_j + 1) \delta_{l_j,r_j} 
- \sqrt{r_j+1} \sqrt{r_j+2} \delta_{l_j,r_j+2} ]. 
\label{p_x_j^2-operator-Hermite-function-matrix-element_2}
\end{eqnarray} 
%}
Therefore, {\sl the dimensionless (nondimensional)} 
quantum Hamiltonian operator $\hat{{\cal H}}_D$ of the quantum harmonic oscillator in $D$ dimensions reads:
%\textcolor{Black}{
\begin{equation}
\hat{{\cal H}}_D = \hat{{\bf p}}^2 + \hat{{\bf x}}^2 = \sum_{j=1}^{D} \hat{{\cal H}}_j , 
\label{H-operator-harmonic-D-definition}
\end{equation}  
%}
acting on the basis function, it produces:
%\textcolor{Black}{
\begin{equation}
\hat{{\cal H}}_D \Psi_{\bf r} = 
{\rm Tr} (2 {\bf r} + {\bf 1}) \Psi_{{\bf r}}, 
\label{H-operator-harmonic-D-Hermite-function-action}
\end{equation}
%}
and its matrix element $({\cal H}_D)_{{\bf l}, {\bf r}}$ is equal:
%\textcolor{Black}{
\begin{equation}
({\cal H}_D)_{{\bf l}, {\bf r}} = 
{\rm Tr} (2 {\bf r} + {\bf 1}) \delta_{{\bf l},{\bf r}}, 
\label{H-operator-harmonic-D-Hermite-function-matrix-element}
\end{equation} 
%}
whereas
%\textcolor{Black}{
\begin{equation}
\epsilon_{\bf r} = 
{\rm Tr} (2 {\bf r} + {\bf 1}) =\sum_{j=1}^{D} \epsilon_{r_{j}}, 
\label{H-operator-harmonic-D-Hermite-function-eigenvalue}
\end{equation} 
%}
is the ${\bf r}$th eigenenergy of $\hat{{\cal H}}_D$.
The eigenenergies are simply the sums of all odd natural numbers, 
and the quantum Hamiltonian is tensor product of diagonal operators in one dimension, 
and its matrix representation is tensor product of diagonal $\infty \times \infty$ matrices
(it is poly-index matrix).
There is no unique straightforward analog of the nearest neighbour spacing $s_r$ in $D$ dimensions.

{\sl Sixthly,} let us perform very difficult task 
consisting of calculating all the absolute moments:
%\textcolor{Black}{
\begin{equation}
(m_{\bf n})_{{\bf l}, {\bf r}}=m_{\bf n} = ({\bf x}^{\bf n})_{{\bf l}, {\bf r}} 
= \langle \Psi_{\bf l} | \hat{{\bf x}}^{\bf n} \Psi_{\bf r} \rangle 
= \prod_{j=1}^{D}\int_{-\infty}^{\infty} \Psi_{l_{j}}^{\star}(x_j) x_j^{n_{j}} \Psi_{r_{j}}(x_j){\rm d}x_j, 
\label{x-operator-nth-abolute-moment-D}
\end{equation} 
%}
of the position operator $\hat{{\bf x}}$ in the Hermite's (Fok's) basis ${\cal B}_D$.
Here, as usual:
%\textcolor{Black}{
\begin{equation}
({\bf x}^{\bf n})_{{\bf l}, {\bf r}} 
= \prod_{j=1}^{D} (x_j^{n_{j}})_{l_j, r_j} . 
\label{x-operator-nth-power-definition-D}
\end{equation} 
%}
One can calculate the lower moments manually, {\sl e. g.}, 
using recurrence relations, matrix algebra, {\sl etc.},
but it is tedious (even for $3 \leq n_j \leq 6$). 
The exact formula for {\sl all} the moments reads:
%\textcolor{Black}{
\begin{equation}
(m_{\bf n})_{{\bf l}, {\bf r}}=m_{\bf n} = ({\bf x}^{\bf n})_{{\bf l}, {\bf r}}
=\prod_{j=1}^{D} (m_{n_{j}})_{l_j, r_j} .
\label{x-operator-nth-abolute-moment-matrix-element-D} 
\end{equation}
%}

{\tt V. Quantum anharmonic oscillator in $D \geq 1$ spatial dimensions.}

{\sl Seventhly,} we are ready to investigate the quantum anharmonic oscillator 
in $D$ spatial dimensions.
Its dimensionless Hamiltonian operator 
$\hat{{\cal H}}_{D, {\rm anharm}}^{{\bf S}}$ reads:
%\textcolor{Black}{
\begin{equation}
\hat{{\cal H}}_{D, {\rm anharm}}^{{\bf S}} 
=\hat{{\cal H}}_D+ \sum_{{\bf s}={\bf 0}}^{{\bf S}} a_{\bf s} \hat{{\bf x}}^{\bf s}
=\hat{{\cal H}}_D+ \sum_{(s_1,...,s_D)=(0,...,0)}^{(S_1,...,S_D)} [a_{(s_1,..,s_D)} \cdot 
\prod_{j=1}^{D}(\hat{x}_j)^{s_j}], 
\label{H-operator-anharmonic-D-S}
\end{equation}
%}
where ${\bf S}$ is a $D$-tuple of degrees of the anharmonicity of the oscillator, 
and the prefactors $a_{\bf s}$ are the strengths of anharmonicity.
The matrix elements of the anharmonic Hamiltonian operator are:
%\textcolor{Black}{
\begin{equation}
({\cal H}_{D, {\rm anharm}}^{{\bf S}})_{{\bf l}, {\bf r}} 
=\epsilon_{\bf r} \delta_{{\bf l}, {\bf r}} 
+ \sum_{{\bf s}={\bf 0}}^{{\bf S}} a_{\bf s} ({\bf{x}}^{\bf s})_{{\bf l}, {\bf r}}, 
\label{H-operator-anharmonic-D-S-matrix-element}
\end{equation}
%}
where the representation of the $D$-dimensional quantum anharmonic oscillator 
in the quantum harmonic oscillator basis ${\cal B}_D$
is mathematically correct, because the basis ${\cal B}_D$ is a complete set, 
and the Hilbert space of the eigenfunctions
of the anharmonic oscillator is isomorphic 
to the Hilbert space ${\cal V}_D$ for the harmonic oscillator, 
provided that
the anharmonic potential in $D$ dimensions:
%\textcolor{Black}{
\begin{equation}
{\cal U}_{D, {\rm anharm}}^{{\bf S}}({\bf x}) = \sum_{{\bf s}={\bf 0}}^{{\bf S}} a_{\bf s} {\bf x}^{\bf s}
=\sum_{(s_1,...,s_D)=(0,...,0)}^{(S_1,...,S_D)} [a_{(s_1,..,s_D)} \cdot 
\prod_{j=1}^{D}(x_j)^{s_j}], 
\label{U-anharmonic-potential-D-S}
\end{equation}
%}
is bounded from below (there are no scattering eigenstates in $D$ dimensions).

{\sl Eighthly,} we repeat that the ``Bohigas conjecture'' also holds for the quantum oscillators 
in $D$ dimensions. 
Having conducted many numerical experiments with different quantum anharmonic oscillators 
(up to the sextic $(D=3)$-dimensional quantum anharmonic oscillators: $S_j=6$) 
we draw conclusion that some of them behave like quantum
integrable systems, the eigenenergies tend to cluster, 
the histograms of nearest neighbour spacing are closer to
the $P_0$ distribution resulting from the Poisson ensemble,
whereas other ones look like quantum chaotic systems, their eigenenergies are subject to repulsion,
the histograms of NNS are closer to the distributions $P_1, P_2, P_4,$ derived from the Gaussian
Random Matrix ensembles.

%}
%end \large

\end{document}